\documentclass[aps,jcp,preprint,groupedaddress]{revtex4}
\usepackage{graphicx}
\usepackage{amsmath}
\usepackage{dcolumn}
\usepackage{bm}

\begin{document}
\title{\textit{Ab initio} potential energy surfaces for 
NH($^3\Sigma^-$) -- NH($^3\Sigma^-$) with analytical long range}

\author{Liesbeth M.~C.~Janssen}
\affiliation{Theoretical Chemistry, Institute for Molecules and Materials (IMM),
Radboud University Nijmegen, Heyendaalseweg 135, 6525 AJ Nijmegen, The Netherlands}
\author{Gerrit C.~Groenenboom}
\affiliation{Theoretical Chemistry, Institute for Molecules and Materials (IMM),
Radboud University Nijmegen, Heyendaalseweg 135, 6525 AJ Nijmegen, The Netherlands}
\author{Ad van der Avoird}
\email[Electronic mail: ]{A.vanderAvoird@theochem.ru.nl}
\affiliation{Theoretical Chemistry, Institute for Molecules and Materials (IMM),
Radboud University Nijmegen, Heyendaalseweg 135, 6525 AJ Nijmegen, The Netherlands}
\author{Piotr S.~\.{Z}uchowski}
\affiliation{Department of Chemistry, Durham University, South Road, DH1 3LE, 
United Kingdom}
\author{Rafa{\l} Podeszwa}
\affiliation{Institute of Chemistry, University of Silesia, Szkolna 9, 40-006
Katowice, Poland}

\date{\today}

\begin{abstract}
We present four-dimensional \textit{ab initio} potential energy surfaces for
the three different spin states of the NH($^3\Sigma^-$) -- NH($^3\Sigma^-$)
complex. The potentials are partially based on the work of Dhont \textit{et
al.}\ [J.\ Chem.\ Phys.\ \textbf{123}, 184302 (2005)].  The surface for the
quintet state is obtained at the RCCSD(T)/aug-cc-pVTZ level of theory and the
energy differences with the singlet and triplet states are calculated at the
CASPT$n$/aug-cc-pVTZ ($n=2,3$) level of theory. The \textit{ab initio}
potentials are fitted to coupled spherical harmonics in the angular
coordinates, and the long range is further expanded as a power series in $1/R$.
The RCCSD(T) potential is corrected for a size-consistency error of about
$0.5\times10^{-6}$ $E_h$ prior to fitting. The long-range coefficients obtained
from the fit are found to be in good agreement with first and second-order
perturbation theory calculations.

\end{abstract}

\maketitle

% 1. INTRODUCTION
\section{Introduction}
The field of cold ($T < 1$ K) and ultracold ($<$ 1 mK) molecules has attracted
great interest in the last few years. The production of such (ultra)cold
species may find important applications in condensed matter physics
\cite{micheli:06}, high precision spectroscopy
\cite{lev:06,bethlem:09,tarbutt:09}, physical chemistry \cite{meerakker:05,
gilijamse:06,gilijamse:07,campbell:08,krems:08}, and quantum computing
\cite{demille:02,andre:06}. There are, in principle, two different strategies
for producing molecular samples at (ultra)low temperatures. In indirect
methods, cold molecules are formed by pairing up atoms that are already cooled
down to the ultracold regime.  Examples of such methods include
photoassociation \cite{jones:06} and Feshbach association \cite{kohler:06}.
Conversely, direct methods such as Stark deceleration \cite{bethlem:03} and
buffer gas cooling \cite{weinstein:98} employ a scheme in which pre-existing
molecules are cooled down from higher temperatures.

One of the most promising candidates for direct-cooling experiments is the NH
radical. NH($X\,^3\Sigma^-$) has a relatively large magnetic moment of 2
$\mu_B$, making it suitable for buffer gas cooling and magnetic trapping
experiments \cite{krems:03a,campbell:07,hummon:08,campbell:08}. Furthermore,
the metastable $a\,^1\Delta$ state of NH, which exhibits a linear Stark effect,
can be efficiently Stark decelerated and trapped in an electrostatic field.
Subsequent excitation of the $A\,^3\Pi \leftarrow a\,^1\Delta$ transition
followed by spontaneous emission to the ground state yields cold
NH($X\,^3\Sigma^-$) molecules, which in turn may be trapped in a magnetic field
\cite{meerakker:01,hoekstra:07}. This scheme also allows for reloading of the
magnetic trap, thus providing a means to increase phase-space density.

At present, direct-cooling methods for NH are limited to temperatures of a few
hundred mK. If the density of trapped molecules is sufficiently high, it may be
possible to reach the ultracold regime by means of evaporative cooling. This
process relies on elastic NH + NH collisions as the trap depth is gradually
reduced. Inelastic spin-changing collisions between trapped NH molecules will
lead to immediate trap loss and are therefore unfavorable. It is generally
accepted that, in order for evaporative cooling to be successful, elastic
collisions should be a few orders of magnitude more efficient than inelastic
transitions
\cite{weinstein:98,balakrishnan:03,campbell:07,krems:03a,cybulski:05}. In the
case of NH($X\,^3\Sigma^-$), the only magnetically trappable state is the
low-field seeking $M_S = 1$ state, with $M_S$ denoting the spin projection
quantum number. A collision complex of two such molecules is in the $M_S = 2$
level of the NH--NH high-spin quintet ($S=2$) state.  Inelastic collisions
between NH molecules may either change the $M_S$ quantum number of the quintet
state, or change the total spin $S$ to produce singlet or triplet complexes.
The $S=0$ and 1 dimer states are chemically reactive \cite{lai:03,poveda:09}
and, although unfavorable for evaporative cooling, could be of interest in the
context of cold controlled chemistry \cite{krems:08}. 

A recent theoretical study by Kajita \cite{kajita:06}, in which only the
electric dipole-induced dipole and magnetic dipole-dipole interactions were
considered, showed that evaporative cooling of NH is likely to be feasible. A
more rigorous quantum calculation of elastic and inelastic cross sections,
however, requires knowledge of the full NH--NH interaction potentials for all
three spin states. In particular the long-range potential, which governs the
dynamics at (ultra)low temperatures, should be described very accurately. For
NH--NH the dominant long-range term is the electrostatic dipole-dipole
interaction, which scales with the intermolecular distance $R$ as $R^{-3}$. If,
however, the molecules are freely rotating, all multipole-multipole terms average
out to zero and the isotropic ($R^{-6}$) dispersion and induction interactions
become important.

Dhont \textit{et al.}\ \cite{dhont:05} have recently constructed
four-dimensional \textit{ab initio} potential energy surfaces for NH--NH which,
in principle, contain all relevant long range contributions. They employed the
partially spin-restricted coupled-cluster method with single and double
excitations and a perturbative treatment of triples [RCCSD(T)]
\cite{knowles:93,knowles:00} to obtain the surface for the NH--NH quintet
state. We found, however, that this surface exhibits erroneous behavior in the
long range due to a lack of size consistency in the open-shell RCCSD(T) method. 
In the
present paper, we report more accurate \textit{ab initio} calculations that are
corrected for this undesirable feature, and which allow for an analytical fit
of the long-range potential. The fit of the short-range potentials is also
improved.
%fit of the singlet and triplet surfaces is also improved.
%are also recalculated with much higher accuracy.

This paper is organized as follows. In Section
\ref{subsec:elecstruc_RCCSD(T)}, we first address the RCCSD(T)
size-consistency problem and present new RCCSD(T) calculations for the long
range of the NH--NH potential.  Long-range perturbation theory calculations are
discussed in Section \ref{subsec:elecstruc_PT}, and new CASPT$n$ ($n = 2,3$)
calculations for the short range of the singlet and triplet potentials are
presented in Section \ref{subsec:elecstruc_CASPTn}. The fit of the different
potentials is described in Section \ref{sec:analytical_repr}, followed by a
discussion of the results in Section \ref{sec:results}. Finally, conclusive
remarks are given in Section \ref{sec:concl}.

% 2. ELECTRONIC STRUCTURE CALCULATIONS
\section{Electronic structure calculations}
\label{sec:elecstruc}

\subsection{RCCSD(T) potential energy surface}
\label{subsec:elecstruc_RCCSD(T)}
The coupled-cluster (CC) approach is one of the most accurate \textit{ab
initio} methods available for calculating potential energy surfaces. This
method requires a single Slater determinant function as the reference state,
which in the case of NH--NH implies that only the high-spin quintet state is
suitable for coupled-cluster calculations. At large intermolecular distances
however, the energy splittings between the three different spin states become
negligible, and thus the CC potential also applies to the singlet and triplet
states at long range. In this section, we will show that the previously
reported NH--NH RCCSD(T) potential \cite{dhont:05} contains a size-consistency
error that becomes apparent at large $R$. We also present new \textit{ab
initio} calculations that are corrected for this defect. The coordinates used
to describe the NH--NH potential energy surfaces are the four intermolecular
Jacobi coordinates ($R$, $\theta_A$, $\theta_B$, $\phi$). The coordinate $R$ is
the length of the intermolecular vector ${\bm R}$ that connects the centers of
mass of monomers $A$ and $B$, $\theta_A$ and $\theta_B$ are the polar angles of
the NH monomer axes relative to ${\bm R}$, and $\phi$ is the dihedral angle
between the planes through ${\bm R}$ and the monomer axes (see also Fig.\ 1 of
Ref.\ \cite{dhont:05}). All interaction potentials are computed using the
supermolecule approach with the counterpoise correction method of Boys and
Bernardi \cite{boys:70}.

\subsubsection{Size consistency}
It is well established that coupled-cluster theory for closed-shell systems is
rigorously size-consistent. For open-shell species, however, where the problem
of nonzero spin arises, this issue is not straightforward. It was demonstrated
in 2006 by Heckert \textit{et al.}\ \cite{heckert:06} that several spin-adapted
CCSD schemes applied to the triplet F($^2P$) -- F($^2P$) system exhibit
size-consistency errors on the order of 10$^{-7}$ -- 10$^{-8}$ $E_h$.  The
reason for this is still unclear, but it has been suggested that the problem
lies in the truncation of the cluster operator \cite{heckert:06}.  Although the
errors are very small, the effect becomes apparent when considering
interactions at low temperatures, where the total energy of the system may be
of a similar order of magnitude (10$^{-7}$ $E_h$ $\approx$ 0.03 K). Hence, a
lack of size consistency imposes a significant limitation on the accuracy of
calculations in the (ultra)cold regime.

When evaluating the NH($^3\Sigma^-$) -- NH($^3\Sigma^-$) quintet potential of
Ref.\ \cite{dhont:05} in more detail, we indeed found that the interaction
energy does not tend to zero at large intermolecular distances. At $R$ =
30$\,$000 $a_0$, the size-consistency error is $-4.8823\times10^{-6}$ $E_h$
calculated at the RCCSD level of theory with the augmented
correlation-consistent polarized valence triple-zeta (aug-cc-pVTZ) basis set
\cite{dunning:89}, and $+0.5129\times10^{-6}$ $E_h$ at the RCCSD(T)/aug-cc-pVTZ
level of theory. It should be noted that these errors are independent of the
relative orientation of the monomers, i.e., the lack of size consistency
affects only the isotropic part of the potential.  The results for other basis
sets are given in Table \ref{tab:sizecon}. It can be seen that the error is
largest at the RCCSD level and increases with the size of the basis set. The
inclusion of triple excitations reduces the error by approximately one order of
magnitude and, for most basis sets, also changes its sign.

Although the problem has not been solved yet, we found that the NH--NH RCCSD(T)
potential can be easily corrected for the lack of size consistency by simply
subtracting the error, calculated at 30\,000 $a_0$, from all \textit{ab initio}
points. We compared these corrected energies with the results obtained from a
spin-unrestricted CCSD(T) [UCCSD(T)] calculation, of which the energies do
converge to zero at long range [i.e.\ UCCSD(T) is size-consistent]. At $R$ =
30.0 $a_0$, the root-mean-square (RMS) difference between the UCCSD(T) and
corrected RCCSD(T) data was calculated to be 9.1$\times10^{-9}$ $E_h$ (0.08\%
of the mean absolute value of the potential) for a grid of 126 \textit{ab
initio} points. Without the size-consistency correction this difference would
be 5.1$\times10^{-7}$ $E_h$ (4.4\%). Thus, the error subtraction at the
RCCSD(T) level leads to significantly better agreement with the size-consistent
UCCSD(T) method.  Similar results were obtained at an intermolecular distance
of 15.0 $a_0$, where the RMS difference between the corrected RCCSD(T) and
UCCSD(T) data is 7.0$\times10^{-8}$ $E_h$ (0.07\% of the mean absolute energy),
as opposed to 5.4$\times10^{-7}$ $E_h$ (0.54\%) without the correction. At even
smaller distances, the size-consistency error will become increasingly
negligible compared to the total interaction energy, thus the correction will
leave the short-range potential essentially unaffected. Based on these
findings, we conclude that subtracting the error from all RCCSD(T) points does
not significantly alter the accuracy of the potential, but does give the
desired asymptotic behavior at long range.

\subsubsection{Long-range RCCSD(T) calculations}
Although the size-consistency correction already constitutes an important
refinement to the RCCSD(T) potential of Ref.\ \cite{dhont:05}, we chose to
improve the long range even further by performing new \textit{ab initio}
calculations. This is motivated by our aim to study collisions in the limit of
zero temperature, for which it is desirable to have the long range in
analytical form.  In order to perform an accurate analytical fit, however, we
found that the long-range \textit{ab initio} energies should be converged to
less than 10$^{-10}$ $E_h$, while the data presented in Ref.\ \cite{dhont:05}
have been converged to only 10$^{-8}$ $E_h$. We therefore recalculated the
points at large $R$ with much tighter convergence thresholds, as low as
10$^{-13}$ $E_h$, to ensure that the fit will not be affected by numerical
noise. The radial grid consisted of 8 points, approximately logarithmically spaced at 8.3,
10.0, 12.0, 14.4, 17.3, 20.8, 25.0, and 30.0 $a_0$. For the angular grid, we
chose an 11-point Gauss-Legendre quadrature grid in ($\theta_A,\theta_B$) and an
11-point Gauss-Chebyshev grid in $\phi$. These are known to be the most
accurate quadratures on their respective domains \cite{stoer:80}. Due to the
symmetry of the complex, only points with $\theta_A + \theta_B \leq \pi$ and $0
\leq \phi \leq \pi$ were required in the calculations \cite{dhont:05}. The
monomers were treated as rigid rotors, with the NH bond length fixed to the
experimental equilibrium value of 1.0362 \AA\ \cite{huber:79}. The RCCSD(T)
energies were computed using the aug-cc-pVTZ basis set, with additional bond
functions located at the midpoint of the intermolecular vector ${\bm R}$
(exponents $s,p$: 0.9, 0.3, and 0.1; $d,f$: 0.6 and 0.2; $g$: 0.3). All
calculations were performed with the MOLPRO package \cite{molpro}. As
explained above, the size-consistency error of 0.51290$\times10^{-6}$ $E_h$ was
subtracted from all RCCSD(T) points to ensure that the long range converges to
zero.

\subsection{Perturbation theory calculations}
\label{subsec:elecstruc_PT}
As an additional test for the accuracy of the RCCSD(T) long-range potential, we
computed the long-range coefficients directly from first and second-order
perturbation theory with the multipole expansion of the interaction operator
(see e.g.\ Ref.\ \cite{stone:96}). The first-order (electrostatic) coefficients
are expressed in terms of the permanent NH multipole moments, while the
second-order (induction and dispersion) coefficients depend also on the static
and dynamic polarizabilities of NH. The permanent multipole moments were
obtained from finite field calculations at the RCCSD(T)/aug-cc-pVTZ level of
theory and from density functional theory (DFT), yielding two different
sets of first-order coefficients. All DFT calculations were performed with the
PBE0 density functional \cite{adamo:99} and the aug-cc-pVQZ basis set. The
Kohn-Sham orbitals were obtained from a spin-restricted calculation using the
DALTON program \cite{dalton2}. The Fermi-Amaldi asymptotic correction
\cite{tozer:98} was employed to improve the description of the NH densities.
The ionization potentials used for this correction were taken from Ref.\
\cite{zuchowski:08a}. For the static and dynamic NH polarizabilities, we
performed spin-restricted time-dependent coupled Kohn-Sham (CKS) calculations
\cite{zuchowski:08a}. Previous studies have shown that CKS methods yield
accurate van der Waals coefficients, comparable to the accuracies obtained with
the best \textit{ab initio} methods, for systems such as He$_2$, Ne$_2$, H$_2$O
dimer \cite{misquitta:05}, and the open-shell O$_2$ dimer \cite{zuchowski:08b}.
The static polarizabilities and dynamic polarizabilities at imaginary
frequencies were obtained with a modified version of the SAPT2008
package \cite{sapt:08}, extended to treat open-shell fragments. Finally, the
second-order long-range coefficients were computed from the DFT multipole
moments and response functions using the POLCOR program
\cite{polcor:92}.

\subsection{CASPT$n$ calculations}
\label{subsec:elecstruc_CASPTn}
As mentioned before, the RCCSD(T) quintet potential can also be used to
describe the singlet and triplet NH--NH states at long range. In the short
range, however, these lower-spin states must be treated with a different
\textit{ab initio} method.  Dhont \textit{et al.}\ \cite{dhont:05} employed the
Complete Active Space with $n$th-order Perturbation Theory (CASPT$n$) method
($n=2,3$) to calculate the energy differences between the quintet state and the
$S$ = 0 and 1 states, and added those to the RCCSD(T) quintet surface to obtain the
singlet and triplet potentials:
\begin{equation}
\label{eq:V_CASPTn}
V^S_n = V^S_{CASPTn} - V^{S=2}_{CASPTn} + V^{S=2}_{RCCSD(T)}.
\end{equation}
When fitting the CASPT$n$ energy splittings, which decay exponentially as a
function of $R$, we found that the convergence thresholds used in Ref.\
\cite{dhont:05} were not sufficiently stringent to reach the same accuracy as
in the long range. Hence, we recalculated the CASPT$n$ energies for all three
spin states with much tighter convergence criteria. The active space consisted
of the four orbitals that are singly occupied in the quintet state. The $g_4$
operator \cite{undoc_g4} was used to obtain size-consistent results, and a
level shift of 0.4 was applied to enforce convergence.  The interaction
energies were computed for $R$ = 4.0, 4.5, 5.0, 5.5, 6.0, 6.5, 7.0, 7.5, 8.0,
8.5, 9.0, 10.0, 12.0, and 14.4 $a_0$, with the energy threshold set to
10$^{-13}$ $E_h$ for the points at 8.0 -- 14.4 $a_0$, 10$^{-12}$ $E_h$ at 7.0
and 7.5 $a_0$, 10$^{-11}$ $E_h$ at 6.0 and 6.5 $a_0$, 10$^{-10}$ $E_h$ at 5.0
and 5.5 $a_0$, 10$^{-9}$ $E_h$ at 4.5 $a_0$, and 10$^{-8}$ $E_h$ at 4.0 $a_0$.
For the angular grid we used the same points as for the long-range RCCSD(T)
calculations, i.e.\ an 11-point Gauss-Legendre quadrature in ($\theta_A$,
$\theta_B$) and an 11-point Gauss-Chebyshev grid in $\phi$. The CASPT$n$
calculations were performed with MOLPRO \cite{molpro} using the
aug-cc-pVTZ basis set supplemented with bond functions. It should be noted that
three points at 4.0 $a_0$ failed to converge due to the strongly repulsive nature
of the potential at small $R$.  
%Nevertheless, we found that the exclusion of
%these points does not significantly alter the results of the fit.

% 3. ANALYTICAL REPRESENTATION
\section{Analytical representation}
\label{sec:analytical_repr}
All three interaction potentials can be represented as follows:
\begin{eqnarray}
\label{eq:LLL}
V(R,\theta_A,\theta_B,\phi) &=& \sum_{L_A,L_B,L}
\upsilon_{L_A,L_B,L}(R) A_{L_A,L_B,L}(\theta_A,\theta_B,\phi) \\
\label{eq:LLM}
&=& \sum_{L_A,L_B,M}
\upsilon_{L_A,L_B,M}(R) A_{L_A,L_B,M}(\theta_A,\theta_B,\phi). 
\end{eqnarray}
The angular functions $A_{L_A,L_B,L}(\theta_A,\theta_B,\phi)$ are defined as
\begin{eqnarray}
\label{eq:A_LLL}
A_{L_A,L_B,L}(\theta_A,\theta_B,\phi) &=& \sum_{M=-\min(L_A,L_B)}^{\min(L_A,L_B)}
\left( \begin{array}{ccc} L_A & L_B & L \\ M & -M & 0 \end{array} \right)
C_{L_A,M}(\theta_A,\phi_A) C_{L_B,-M}(\theta_B,\phi_B), \nonumber \\
&=& \sum_{M=0}^{\min(L_A,L_B)} (-1)^{M}
\left( \begin{array}{ccc} L_A & L_B & L \\ M & -M & 0 \end{array} \right)
A_{L_A,L_B,M}(\theta_A,\theta_B,\phi),
\end{eqnarray}
where $C_{L,M}(\theta,\phi)$ are Racah-normalized spherical harmonics and
$\phi=\phi_A-\phi_B$ is the difference between the azimuthal angles of monomers
$A$ and $B$. The factor in brackets denotes a Wigner three-$j$ symbol.  The
`primitive' angular functions $A_{L_A,L_B,M}(\theta_A,\theta_B,\phi)$ are given
by
\begin{equation}
\label{eq:A_LLM}
A_{L_A,L_B,M}(\theta_A,\theta_B,\phi) = P_{L_A,M}(\cos{\theta_A})
P_{L_B,M}(\cos{\theta_B})\cos{M\phi},
\end{equation}
where $P_{L,M}(\cos{\theta})$ are Schmidt semi-normalized associated
Legendre functions defined in Ref.\ \cite{dhont:05}.  The $R$-dependent
expansion coefficients are related to each other as \cite{dhont:05}
\begin{equation}
\label{eq:vLLL}
\upsilon_{L_A,L_B,L}(R) = (2L+1) \sum_{M=0}^{\min(L_A,L_B)} (-1)^M (2-\delta_{M0})
\left( \begin{array}{ccc} L_A & L_B & L \\ M & -M & 0 \end{array} \right)
\upsilon_{L_A,L_B,M}(R).
\end{equation}

\subsection{Long-range potential}
\label{sec:lr}
For the analytical long-range interaction, we use Eq.\ (\ref{eq:LLL}) and
further expand the $\upsilon_{L_A,L_B,L}(R)$ coefficients in a power series in
$1/R$:
\begin{equation}
\label{eq:Vlr}
\upsilon_{L_A,L_B,L}(R) = \sum_{n} \frac{-C_{L_A,L_B,L,n}}{R^n}.
\end{equation}
Our choice of an 11-point Gauss-Legendre quadrature in ($\theta_A$, $\theta_B$)
and an 11-point Gauss-Chebyshev quadrature in $\phi$ ensures that the angular
$A_{L_A,L_B,L}$ functions, when evaluated on the quadrature grid with the
appropriate weights, are mutually orthogonal for all values of $L_A$ and $L_B$
up to 10 inclusive.  Thus, we can perform the analytical fit in $R$ [Eq.\
(\ref{eq:Vlr})] for each ($L_A,L_B,L$) term separately. The values of $n$
follow from a consideration of the possible first-order (electrostatic) and
second-order (induction/dispersion) contributions (see e.g.\ Ref.\
\cite{avoird:80} for details). For the electrostatic terms, we have $L_A + L_B
= L$ and $n = L_A + L_B + 1$, with $L_A \geq 1$ and $L_B \geq 1$.  The minimum
value of 1 comes from the fact that the lowest nonvanishing permanent multipole
moment of NH is the dipole. In the case of induction and dispersion
interactions, $L_A$ and $L_B$ correspond to the order of two coupled multipole
moments on monomers $A$ and $B$, respectively.  That is, $L_A =
|l_A-l_A'|,\hdots,l_A+l_A'$ and $L_B = |l_B-l_B'|, \hdots,l_B+l_B'$, where
$l_A$, $l_A'$, $l_B$, and $l_B'$ denote the orders of the uncoupled monomer
multipole moments. $L_A$ and $L_B$ are in turn coupled to all possible $L$
values, and for each ($L_A$,$L_B$,$L$) term we have $n=l_A+l_A'+l_B+l_B'+2$.
Finally, due to the inversion symmetry of the total system, it can be shown
that $L_A+L_B+L$ is even, and since each monomer is a linear $\Sigma$ state
molecule, $l_A+l_A'+L_A$ and $l_B+l_B'+L_B$ must also be even \cite{avoird:80}.

The $C_{L_A,L_B,L,n}$ fit coefficients of Eq.\ (\ref{eq:Vlr}) were calculated
as follows. For each set of ($L_A,L_B,L$) values, we first computed the lowest
possible values of $n$ in both first and second order. Since our long-range
\textit{ab initio} calculations were performed on a grid of eight $R$ points,
we could include a maximum of eight $R^{-n}$ functions in the fit.  We then
fitted the size-consistency corrected RCCSD(T) data to the expansion of Eq.\
(\ref{eq:LLL}), and subsequently fitted each $\upsilon_{L_A,L_B,L}$ expansion
coefficient in terms of $R^{-n}$ functions [Eq.\ (\ref{eq:Vlr})].  Note that
the fit of Eq.\ (\ref{eq:LLL}) is mathematically equivalent to evaluating the
overlap integral between the angular functions and
$V(R,\theta_A,\theta_B,\phi)$ by Gauss-Legendre quadrature.
The fit was done using a linear least-squares procedure in which the
\textit{ab initio} points were weighted with the appropriate quadrature weights
and a factor of $R^3$.  The $R$-dependent factor is chosen because the leading
dipole-dipole interaction decays as $R^{-3}$.

In principle, our long-range expansion is valid for all terms up to
$L_A=L_B=10$, with eight possible values of $n$ for each ($L_A,L_B,L$) term.
However, the inclusion of high powers of $1/R$ may lead to unphysical results
even for the low-$n$ coefficients, which are considered the most important.
Thus, we must carefully choose which $R^{-n}
A_{L_A,L_B,L}(\theta_A,\theta_B,\phi)$ functions to include in the fit.  After
extensive testing, we found that the best analytical fit is obtained for $n
\leq 14$. This result is based on a thorough examination of both the stability
of the fit, i.e.\ how much the $C_{L_A,L_B,L,n}$ coefficients vary when adding
more $R^{-n}$ functions, and the RMS error in the data points. The final fit
gave a RMS error of 4.6$\times10^{-8}$ $E_h$ (0.03\%) for a total of 10648
\textit{ab initio} points. The RMS difference between the analytical potential
and the size-consistency corrected long-range points of Ref.\ \cite{dhont:05},
which served as test points, was 4.8$\times10^{-7}$ $E_h$ (0.24\%). Note that
the latter error is, in part, due to the weaker convergence thresholds used in
the calculations of Ref.\ \cite{dhont:05}. The $C_{L_A,L_B,L,n}$ fit coefficients
are available through EPAPS \cite{janssen_epaps:09}.

\subsection{Short-range $S = 2$ potential}
For the short range of the quintet surface, we used the size-consistency
corrected RCCSD(T) data of Dhont \textit{et al.}\ \cite{dhont:05}, calculated
at $R$ values from 4.0 to 16.0 $a_0$.  The angular grid consisted of 11 points
in $\theta_A$ and $\theta_B$, ranging from 0$^{\circ}$ to 180$^{\circ}$ in
steps of 20$^{\circ}$ with an additional point at 90$^{\circ}$.  The grid in
$\phi$ ranged from 0$^{\circ}$ to 180$^{\circ}$ in steps of 22.5$^{\circ}$.
The short-range potential was first expanded in terms of
$A_{L_A,L_B,M}(\theta_A,\theta_B,\phi)$ functions [Eq.\ (\ref{eq:LLM})] and
then transformed to Eq.\ (\ref{eq:LLL}).  Instead of using the two-step
spline-based approach described in Ref.\ \cite{dhont:05}, we employed a
weighted least squares fitting procedure to determine the
$\upsilon_{L_A,L_B,M}(R)$ coefficients for each $R$. In order to perform the
fit, we first calculated optimal quadrature weights for the grid points in
($\theta_A,\theta_B$), of which the details are given in the Appendix.  We then
attempted to fit the RCCSD(T) points by an expansion in terms of
$A_{L_A,L_B,M}(\theta_A,\theta_B,\phi)$ functions, weighting each point with
the appropriate quadrature weights. High-energy points ($>$ 0.1 $E_h$), which
are not of practical importance in bound-state and scattering calculations,
were excluded from the fit. It was found, however, that the least squares
problem of Eq.\ (\ref{eq:LLM}) is ill-conditioned for max($L_A,L_B$) $\geq$ 9
due to both the choice of grid points (the angle $\phi$ is undefined if
$\theta_A$ or $\theta_B$ equals 0$^{\circ}$ or 180$^{\circ}$) and the omission
of points at high energies. We therefore employed a modified fitting scheme to
regularize the least squares problem such that all functions up to $L_A=L_B=10$
and $M=8$ could be included. This was done by means of a Tikhonov
regularization method \cite{tikhonov:77} in which the term $\sum_{L_A,L_B,M} |
\alpha(L_A^2 + L_B^2) \upsilon_{L_A,L_B,M}(R) |^2$ was added to the residual.
The factor of $\alpha(L_A^2 + L_B^2)$, with $\alpha = 2\times10^{-4}$, ensures
that strong oscillations (associated with large $L_A$ and $L_B$) are damped out
in the fit. The resulting $\upsilon_{L_A,L_B,M}(R)$ fit coefficients were then
transformed to $\upsilon_{L_A,L_B,L}(R)$ coefficients using Eq.\
(\ref{eq:vLLL}). Overall, this fitting procedure gave a RMS error of
9.8$\times10^{-6}$ $E_h$ (0.21\%) based on 21275 \textit{ab initio} points.
The $\upsilon_{L_A,L_B,L}(R)$ coefficients can be retrieved via the
EPAPS system \cite{janssen_epaps:09}.

The $\upsilon_{L_A,L_B,L}(R)$ expansion coefficients were interpolated in $R$
using the reproducing kernel Hilbert space (RKHS) method with the reproducing
kernel for distancelike variables \cite{ho:96,ho:00}. The RKHS parameter $m$,
which determines the power with which the interpolated function decays between
the grid points, was set to the leading power in $1/R$ for each ($L_A,L_B,L$)
term. For instance, the $\upsilon_{112}(R)$ coefficient containing the
electrostatic dipole-dipole interaction was interpolated with $m=3$, the
isotropic $\upsilon_{000}(R)$ term was interpolated with $m=6$, and so on. In
all cases, the RKHS smoothness parameter was set to 2. 

Finally, we matched the short-range and long-range expansions of the RCCSD(T)
quintet potential using an $R$-dependent switching function $f(R)$ that changes
smoothly from 0 to 1 on the interval $a<R<b$:
\begin{equation}
\label{eq:switch}
f(R) = \left\{ \begin{array}{ll} 0 & \mbox{if $R \leq a$} \\
                                 1 & \mbox{if $R \geq b$} \\
        \frac{1}{2} + \frac{1}{4} \sin{\frac{\pi x}{2}}
        \left(3 - \sin^2{\frac{\pi x}{2}} \right)
                       & \mbox{otherwise,} \end{array} \right.
\end{equation}
with $x = \frac{(R-b)+(R-a)}{b-a}$. The function is such that the first three
derivatives at $R=a$ and $R=b$ are zero.  We used Eq.\ (\ref{eq:switch}) to
switch the potential between $a=8$ and $b=12$ $a_0$. The total $S = 2$
potential energy surface may now be expressed as follows:
\begin{equation}
\label{eq:Vtot}
V(R,\theta_A,\theta_B,\phi) = [1-f(R)] V_{sr}(R,\theta_A,\theta_B,\phi)
                              + f(R) V_{lr}(R,\theta_A,\theta_B,\phi),
\end{equation}
where $V_{sr}$ refers to the short-range expansion of Eq.\ (\ref{eq:LLL}) and
$V_{lr}$ to the long-range expansion of Eqs.\ (\ref{eq:LLL}) and
(\ref{eq:Vlr}).

\subsection{Short-range $S = 0, 1$ potentials} As already mentioned in Section
\ref{sec:elecstruc}, the singlet and triplet potentials were obtained from the
quintet RCCSD(T) potential by adding the energy differences at the CASPT2 or
CASPT3 level of theory. We fitted these exchange splittings ($V^S_{CASPTn} -
V^{S=2}_{CASPTn}$) directly in terms of $A_{L_A,L_B,L}(\theta_A,\theta_B,\phi)$
functions, weighting each point with the corresponding Gauss-Legendre and
Gauss-Chebyshev quadrature weights.  In all cases, the fit error was largest at
4.0 $a_0$ and rapidly decreased as a function of $R$. For instance, the RMS
errors for the singlet-quintet CASPT2 and CASPT3 splittings were
1.3$\times$10$^{-3}$ $E_h$ (4.6\%) and 1.2$\times$10$^{-3}$ $E_h$ (4.7\%) at
4.0 $a_0$, 1.1$\times$10$^{-5}$ $E_h$ (0.10\%) and 7.8$\times$10$^{-6}$ $E_h$
(0.09\%) at the neighboring grid point of 4.5 $a_0$, and 2.3$\times$10$^{-8}$
$E_h$ (0.007\%) and 1.9$\times$10$^{-8}$ $E_h$ (0.007\%) near the van der Waals
minimum at 6.5 $a_0$. For the triplet-quintet CASPT2 and CASPT3 exchange
splittings, the RMS errors were 6.9$\times$10$^{-4}$ $E_h$ (3.2\%) and
7.9$\times$10$^{-3}$ $E_h$ (4.4\%) at 4.0 $a_0$, 4.3$\times$10$^{-6}$ $E_h$
(0.06\%) and 5.1$\times$10$^{-6}$ $E_h$ (0.08\%) at 4.5 $a_0$, and
2.1$\times$10$^{-8}$ $E_h$ (0.01\%) and 6.0$\times$10$^{-8}$ $E_h$ (0.03\%) at
6.5 $a_0$. All errors were calculated from 1331 \textit{ab initio} points per
$R$ value, with the exception of $R$ = 4.0 $a_0$, where three points failed to
converge. The $\upsilon_{L_A,L_B,L}(R)$ fit coefficients for the CASPT$n$
energy splittings are available through EPAPS \cite{janssen_epaps:09}.

The $\upsilon_{L_A,L_B,L}(R)$ coefficients were interpolated in $R$ using
the RKHS method. For all ($L_A,L_B,L$) terms we set the RKHS parameter $m$ to
14 and the smoothness parameter to 2. The value of $m=14$ ensures that all
coefficients decay as $R^{-15}$ beyond the outermost grid point, thus decaying
faster than any of the long-range terms included in the fit of Eq.\
(\ref{eq:Vlr}). In addition, we found that the interpolation with $m=14$ gives
the smallest RMS error in the \textit{ab initio} points of Ref.\
\cite{dhont:05}. The expanded CASPT$n$ splittings were added to the RCCSD(T)
potential of Eq.\ (\ref{eq:Vtot}) to obtain the final singlet and triplet
potential energy surfaces.

% 4. RESULTS AND DISCUSSION
\section{Results and Discussion}
\label{sec:results}

The main features of the singlet, triplet, and quintet potentials have already
been described in Ref.\ \cite{dhont:05}, and therefore we only briefly mention
them here. Our $S=2$ potential is characterized by a van der Waals minimum at
$R_e$ = 6.61 $a_0$ with a well depth of $D_e$ = $-$675 cm$^{-1}$.  It should be
noted that Dhont \textit{et al.}\ \cite{dhont:05} reported a slightly different
$R_e$ value of 6.60 $a_0$. The minimum corresponds to a linear geometry
($\theta_A = \theta_B = \phi = 0^{\circ}$) in which the two NH dipoles are
aligned. \.{Z}uchowski \textit{et al.}\ \cite{zuchowski:08a} have recently
shown that $D_e$ changes to $-$693 cm$^{-1}$ if the aug-cc-pVQZ basis is used
and the RCCSD(T) calculations are performed without the frozen-core
approximation. They also demonstrated from symmetry-adapted perturbation theory
(SAPT) calculations that the main contributions to $D_e$ are the electrostatic
($-$899 cm$^{-1}$) and dispersion ($-$432 cm$^{-1}$) interactions. The total
SAPT exchange-repulsion energy at the minimum was found to be 874 cm$^{-1}$
\cite{zuchowski:08a}.

The $V^{S=0}_2$ ($V^{S=0}_3$) and $V^{S=1}_2$ ($V^{S=1}_3$) surfaces also
exhibit a van der Waals minimum at $\theta_A = \theta_B = \phi = 0^{\circ}$,
located at $R_e$ = 6.50 (6.51) and 6.54 (6.55) $a_0$, respectively. These
distances are 0.01 -- 0.02 $a_0$ different from the $R_e$ values reported by
Dhont \textit{et al.}\ \cite{dhont:05}.  Furthermore, the singlet and triplet
dimers may form the chemically stable N$_2$H$_2$ molecule, which is reflected
in the strongly attractive nature of these potentials at short intermolecular
separations. The most favorable geometries for the $S = 0$ and 1 states at
short distances are found near $\theta_A = \theta_B = 90^{\circ}$.

\subsection{Long-range potential}

Before discussing the analytical fit results, we first address the
size-consistency problem occurring at the RCCSD and RCCSD(T) levels of theory.
Figure \ref{fig:v000} shows the isotropic part of the quintet potential,
$\upsilon_{000}(R)$, between $R$ = 15 and 30 $a_0$. The lack of size
consistency is most apparent at the RCCSD level, giving rise to an error of
$-$1.07 cm$^{-1}$ at long range. The inclusion of triple excitations reduces
the problem significantly, but in fact overcompensates for the RCCSD error by
+0.11 cm$^{-1}$. The uncorrected isotropic RCCSD and RCCSD(T) potentials cross
at $R \approx 11$ $a_0$.  After subtracting the size-consistency errors from
all \textit{ab initio} points, both the RCCSD and RCCSD(T) potentials smoothly
converge to zero at long range. It can also be seen that these corrected data
are in very good agreement with the corresponding spin-unrestricted CC results
at $R$ = 15 and 30 $a_0$.

The main fit results for the (size-consistency corrected) RCCSD(T) long-range
potential are presented in Table \ref{tab:Cn}. A total number of 588
$C_{L_A,L_B,L,n}$ coefficients was included in the long-range fit ($L_A, L_B
\leq 10$ and $n \leq 14$), but here we list only the most important terms.
Table \ref{tab:Cn} also shows the results obtained from first and second-order
perturbation theory (PT).  It can be seen that the fitted electrostatic terms
agree very well with the PT coefficients, in particular with the data
calculated at the PT--RCCSD(T) level of theory. For the induction and
dispersion terms we find some significant discrepancies, but the most important
second-order fit coefficients (those with $n = 6$) show satisfactory agreement
with PT--DFT. It should be noted that, for the fitted coefficients, no
distinction can be made between induction and dispersion contributions. For the
isotropic $C_{0,0,0,6}$ term, the PT-DFT calculations give a dispersion
coefficient of 39.86 a.u.\ and an induction term of 6.99 a.u.

As an indication of the relative importance of the different $C_{L_A,L_B,L,n}$
coefficients, we explicitly give their contributions to the potential at $R$ =
30 $a_0$ (see Table \ref{tab:Cn}). These contributions, $V_{L_A,L_B,L,n}(R)$,
were calculated as follows:
\begin{equation}
V_{L_A,L_B,L,n}(R) = N_{L_A,L_B,L} \frac{|C_{L_A,L_B,L,n}^{fit}|}{R^{n}},
\end{equation}
where $N_{L_A,L_B,L}$ = $[4\pi/(2L_A+1)(2L_B+1)(2L+1)]^{1/2}$ is the norm of
the angular $A_{L_A,L_B,L}(\theta_A,\theta_B,\phi)$ functions.  It is clear
that the $n = 3$ dipole-dipole interaction dominates the potential by at least
one order of magnitude, followed by the electrostatic dipole-quadrupole term.
The main second order term is the isotropic $n = 6$ interaction, which, at 30
$a_0$, is still larger than the electrostatic $n = 5$ terms. The fact that the
fitted $C_{1,1,2,3}$ and $C_{0,0,0,6}$ coefficients give the largest
contributions in first and second order, respectively, indicates that the fit
is not only numerical, but also physically meaningful.  Thus, we may safely
extrapolate the potential from 30 $a_0$ to larger $R$ values.

Figure \ref{fig:Vfitlr} shows the $R$-dependence of the fitted RCCSD(T)
potential for two specific orientations ($\theta_A,\theta_B,\phi$). For the
linear geometry, with $\theta_A = \theta_B = \phi$ = $0^{\circ}$, the leading
dipole-dipole interaction is maximally attractive, while for $\theta_A =
\theta_B = \phi$ = $90^{\circ}$ the dipole-dipole term is zero.  It can be seen
that the $C_{1,1,2,3}$ coefficient dominates the long-range potential beyond $R
\approx 12$ $a_0$.  Figure \ref{fig:Vfitlr} also compares the total long-range
expansion with the \textit{ab initio} data, illustrating the region of validity
of Eq.\ (\ref{eq:Vlr}).  It should be noted that, on the scale of the figure,
the short-range expansion of Eq.\ (\ref{eq:LLL}) is indistinguishable from the
total fitted potential of Eq.\ (\ref{eq:Vtot}), and thus the short-range
expansion is not explicitly shown.  The long-range fit is very accurate for
intermolecular distances larger than 8 $a_0$, which suggests that short-range
(exchange and charge penetration) effects are only significant for $R \leq 8$
$a_0$. This also justifies our choice of switching the potential from the
short-range to the long-range expansion between 8 and 12 $a_0$.

\subsection{Short-range potentials}
Although the $S = 0$, 1, and 2 potentials obtained in this work are very
similar to those reported by Dhont \textit{et al.}\ \cite{dhont:05}, there are
some notable differences at very short intermolecular distances. The
differences are most pronounced at $R$ = 4.0 $a_0$, where the potentials
exhibit the highest anisotropy.  Figure \ref{fig:Vfit_q4a0} compares the two
fit results for the quintet state as a function of $\theta_A$ and $\theta_B$,
with $R$ = 4.0 $a_0$ and $\phi = 0^{\circ}$.  Note that both surfaces were
obtained from the same set of \textit{ab initio} data.  The fit of Ref.\
\cite{dhont:05} shows more oscillatory behavior than our present result, in
particular near ($\theta_A,\theta_B$) = ($180^{\circ},150^{\circ}$) and
($30^{\circ},0^{\circ}$). Furthermore, the potential of Dhont \textit{et al.}\
has a local maximum around ($150^{\circ},30^{\circ}$) that is clearly
unphysical in nature.  Similar patterns are found for the triplet and singlet
states, as can be seen in Figs.\ \ref{fig:Vfit_t4a0} and \ref{fig:Vfit_s4a0}.
The $S = 0$ and 1 potentials of Ref.\ \cite{dhont:05} exhibit more pronounced
oscillations and local maxima, indicating more unphysical behavior. We therefore
conclude that, in addition to the more accurate long-range potential, the fit of 
the short-range NH--NH potentials is also improved in the present work.

% 5. CONCLUSIONS AND OUTLOOK
\section{Conclusions and Outlook}
\label{sec:concl}
We have constructed four-dimensional potential energy surfaces for the singlet,
triplet, and quintet states of NH($^3\Sigma^-$) -- NH($^3\Sigma^-$) based on
high-level \textit{ab initio} calculations. All potentials were fitted in terms
of coupled spherical harmonics in the angular coordinates, and the long range
was further expanded as a power series in $1/R$. Prior to fitting, the
\textit{ab initio} data were corrected for a size-consistency error of
$0.5\times10^{-6}$ $E_h$ occurring at the RCCSD(T) level of theory. The fitted
long-range coefficients were found to be in good agreement with the results
obtained from first and second-order perturbation theory. 

Future work is planned to study the evaporative cooling process of NH, which
requires knowledge of elastic and inelastic cross sections at (ultra)low
temperatures.  Rate constants and cross sections for (cold) reactive NH + NH
collisions will also be calculated.  Finally, we aim to explore the
possibilities of cold controlled chemistry by investigating the influence of
external fields.

% 6. ACKNOWLEDGMENTS
\begin{acknowledgments}
We gratefully acknowledge Professor Robert Moszy\'{n}ski and Professor
Hans-Joachim Werner for useful discussions on the RCCSD(T) size-consistency
problem. LMCJ and GCG thank the Council for Chemical Sciences of the
Netherlands Organization for Scientific Research (CW-NWO) for financial
support. PSZ acknowledges EPSRC for funding the collaborative project CoPoMol
under the ESF EUROCORES programme EuroQUAM.

\end{acknowledgments}

% APPENDIX
\appendix*
\section{}
\label{sec:appendix}
In this Appendix, we describe how we optimized the quadrature weights
$w_i$ for the integration of Legendre polynomials $P_l(x)$ on a given grid of
mutually distinct points $x_i$ ($i=1,\hdots,n$):
\begin{equation}
\int_{-1}^{1} P_l(x) dx = 2 \delta_{l,0} 
\approx \sum_{i=1}^{n} w_i P_l(x_i).
\end{equation}
We define the optimization as a minimization of the sum of square residuals
$|{\bm r}|$:
\begin{equation}
|{\bm r}| = |{\bm A} {\bm w} - {\bm b}|,
\end{equation}
where ${\bm A}$ is an $(l_{max}+1) \times n$ matrix with elements $A_{li} =
P_l(x_i)$ ($l = 0,\hdots,l_{max}$), ${\bm w}$ is a vector of length $n$
containing the quadrature weights $w_i$, and ${\bm b}$ is a vector of length
$l_{max}+1$ with elements $b_l = 2\delta_{l,0}$. 
In the case of an $n$-point Gauss-Legendre quadrature, $x_i$ and $w_i$ are
chosen in such a way that the integration is exact, i.e., $|{\bm r}| = 0$,
for all polynomials up to degree $l_{max} = 2n-1$. For arbitrary, mutually
distinct points $x_i$, we may calculate the weights as ${\bm w} = {\bm A^{-1}}
{\bm b}$, since ${\bm A}$ is regular for $l_{max} = n-1$ (see p.\ 145 of 
Ref.\ \cite{stoer:80}). This results in a quadrature that is exact up to
(at least) degree $n-1$. Instead of using a quadrature that is exact for
$l_{max} = n-1$ and most likely unsuitable for higher degree polynomials,
we choose a compromise quadrature that is reasonable for $l_{max} > n-1$ at the
expense of not being exact for lower degree polynomials. This may be
achieved by linear least squares minimization of $|{\bm r}|$. However,
we prefer to use a quadrature that is exact for constant functions ($l=0$), which
requires a minimization of $|{\bm r}|$ with the constraint that
$\sum_{i=1}^{n} w_i = 2$. For this purpose we take 
\begin{equation}
\label{eq:w}
{\bm w} = {\bm w_0} + {\bm w_{\perp}},
\end{equation}
with $({\bm w_0})_i = 2/n$ for all $i = 1,\hdots,n$ and 
$\sum_{i=1}^{n} ({\bm w_{\perp}})_i = 0$.
This may be rewritten as ${\bm w_0^T}{\bm w_{\perp}} = 0$, with ${\bm w_0^T}$
denoting the transpose of ${\bm w_0}$. We can now expand
${\bm w_{\perp}}$ in an orthogonal basis $\{{\bm q_i}, i=2,\hdots,n\}$
of vectors ${\bm q_i}$ that are perpendicular to ${\bm w_0}$:
\begin{equation}
\label{eq:w_perp}
{\bm w_{\perp}} = \sum_{i=2}^{n} {\bm q_i} c_i = \tilde{{\bm Q}} {\bm c}.
\end{equation}
We observe that the first row of the matrix ${\bm A}$ is proportional to
${\bm w_0}$, and thus the vectors ${\bm q_i}$ can be generated by
Gram-Schmidt QR-factorization of ${\bm A^{T}}$:
\begin{equation}
{\bm A^T} = {\bm Q} {\bm R}.
\end{equation}
Here, ${\bm Q}$ is an $n \times n$ orthonormal matrix and ${\bm R}$ is an $n
\times (l_{max}+1)$ upper triangular matrix. The columns $i = 2,\hdots,n$
of ${\bm Q}$ form the matrix $\tilde{{\bm Q}}$ of Eq.\ (\ref{eq:w_perp}).  In
order to find the expansion coefficients ${\bm c}$, we now remove the first row
of ${\bm A}$ and the first element of ${\bm b}$, yielding the ($l_{max} \times
n$) matrix $\tilde{{\bm A}}$ and the null vector $\tilde{{\bm b}}$ of length
$l_{max}$, respectively, and define the residual $\tilde{{\bm r}} = \tilde{{\bm
A}} {\bm w}$. Substitution of Eq.\ (\ref{eq:w}) gives
\begin{equation}
|\tilde{{\bm r}}| = |\tilde{{\bm A}} {\bm w_0} + 
                     \tilde{{\bm A}} \tilde{{\bm Q}} {\bm c}|,
\end{equation}
which can be minimized in a standard least squares procedure to obtain the
expansion coefficients ${\bm c}$. Finally, substitution of Eq.\
(\ref{eq:w_perp}) into (\ref{eq:w}) gives the total optimal quadrature weights.
In the present work, we have employed this method to generate optimal weights
for the short-range quintet potential with $n=11$ and $l_{max}=16$.

% \bibliography{vanderwaals}

\clearpage
\begin{table}
\caption{Size-consistency errors ($\Delta E$) for the NH--NH system at the
RCCSD and RCCSD(T) levels of theory. The basis sets correspond to the
(aug)-cc-pV$n$Z ($n$ = double, triple, quadruple, quintuple) sets of Dunning
\cite{dunning:89}.  The errors are calculated as the difference between the
energy of the separate monomers and the energy of the supersystem NH$\cdots$NH
at an intermolecular distance of 30$\,$000 $a_0$. All values are in 10$^{-6}$
$E_h$.
\label{tab:sizecon}}
\centering
\begin{ruledtabular}
\begin{tabular}{l c c}
Basis set & \multicolumn{1}{l}{$\Delta E$ RCCSD} & \multicolumn{1}{l}{$\Delta
E$ RCCSD(T)} \\
\hline
cc-pVDZ      & $-3.15067$ & $-0.50946$  \\
cc-pVTZ      & $-4.25041$ & $-0.01069$  \\
cc-pVQZ      & $-4.70853$ & $ 0.36976$  \\
cc-pV5Z      & $-4.92130$ & $ 0.62672$  \\
\\
aug-cc-pVDZ  & $-4.04159$ & $ 0.01944$  \\
aug-cc-pVTZ  & $-4.88230$ & $ 0.51290$  \\
aug-cc-pVQZ  & $-5.01375$ & $ 0.68981$  \\
aug-cc-pV5Z  & $-5.03493$ & $ 0.75827$  \\
\end{tabular}
\end{ruledtabular}
\end{table}

\clearpage
\begin{table}
\caption{Most important long-range coefficients obtained from the fit and from
perturbation theory, and their contributions at 30 $a_0$.  The order of
importance is based on the value of $n$, and for each $n$ only the four largest
terms are given. Terms labeled with an asterisk are first-order (electrostatic)
interactions.  All values are in atomic units. Numbers in parentheses denote
powers of 10.
\label{tab:Cn}}
\centering
\begin{ruledtabular}
\begin{tabular}{l l l l r  r r r}
$L_A$ & $L_B$ & $L$ & $n$ & $C_{L_A,L_B,L,n}^{fit}$ &
$C_{L_A,L_B,L,n}^{PT-RCCSD(T)}$ & $C_{L_A,L_B,L,n}^{PT-DFT}$ &
$V_{L_A,L_B,L,n}$(30 $a_0$) \\
\hline
  1  &  1  &  2  &  3*   &   1.9697(+0)  &  1.9715(+0) &   2.0127(+0)    &   3.8551(-05)\\
  1  &  2  &  3  &  4*   &  -2.8394(+0)  & -2.8597(+0) &  -3.0642(+0)    &   1.2127(-06)\\
  1  &  3  &  4  &  5*   &   1.6637(+1)  &  1.6761(+1) &   1.7103(+1)    &   1.7654(-07)\\
  2  &  2  &  4  &  5*   &  -5.6953(+0)  & -5.4312(+0) &  -6.1080(+0)    &   5.5389(-08)\\
  0  &  0  &  0  &  6    &   4.7270(+1)  &             &   4.6852(+1)    &   2.2986(-07)\\
  1  &  4  &  5  &  6*   &  -5.4131(+1)  & -5.5049(+1) &  -5.7422(+1)    &   1.5274(-08)\\
  0  &  2  &  2  &  6    &   1.2657(+1)  &             &   1.5681(+1)    &   1.2309(-08)\\
  2  &  3  &  5  &  6*   &   3.6904(+1)  &  3.9347(+1) &   4.2140(+1)    &   9.1458(-09)\\
%  2  &  2  &  4  &  6    &   2.4781(+1)  &             &   1.1712(+1)    &   8.0335(-09)\\
  0  &  1  &  1  &  7    &  -1.8433(+2)  &             &  -8.2153(+1)    &   9.9596(-09)\\
  1  &  2  &  3  &  7    &  -3.4979(+2)  &             &   1.5651(+1)    &   5.5331(-09)\\
  0  &  3  &  3  &  7    &  -1.0784(+2)  &             &  -7.9522(+1)    &   2.4971(-09)\\
  3  &  3  &  6  &  7*   &   3.1701(+2)  &  3.3946(+2) &   3.4622(+2)    &   2.0359(-09)\\
%  1  &  2  &  1  &  7    &   6.0628(+1)  &             &  -3.1371(+0)    &   1.4650(-09)\\
  0  &  0  &  0  &  8    &   9.2546(+2)  &             &   1.1077(+3)    &   5.0003(-09)\\
  0  &  2  &  2  &  8    &   3.9371(+3)  &             &   1.4208(+3)    &   4.2544(-09)\\
  1  &  1  &  2  &  8    &   4.5792(+3)  &             &  -1.0618(+2)    &   3.6882(-09)\\
  2  &  2  &  4  &  8    &  -4.4826(+3)  &             &   6.2384(+2)    &   1.6146(-09)\\
%  1  &  3  &  4  &  8    &   3.3905(+3)  &             &  -1.1118(+2)    &   1.3325(-09)\\
  0  &  1  &  1  &  9    &   3.0500(+4)  &             &  -3.1644(+3)    &   1.8310(-09)\\
  1  &  2  &  3  &  9    &   9.9936(+4)  &             &   2.2929(+3)    &   1.7565(-09)\\
  1  &  2  &  1  &  9    &  -2.3093(+4)  &             &  -6.1280(+2)    &   6.2000(-10)\\
  0  &  3  &  3  &  9    &   9.0400(+3)  &             &  -5.6295(+3)    &   2.3259(-10)\\
%  1  &  4  &  5  &  9    &  -1.3535(+4)  &             &   3.7372(+2)    &   1.4145(-10)\\
%  1  &  1  &  2  & 10    &  -1.8285(+6)  &             &  -6.6595(+3)    &   1.6364(-09)\\
%  0  &  2  &  2  & 10    &  -5.9888(+5)  &             &   4.8649(+4)    &   7.1906(-10)\\
%  0  &  0  &  0  & 10    &   9.4520(+4)  &             &   2.7609(+4)    &   5.6743(-10)\\
%  1  &  3  &  4  & 10    &  -1.1187(+6)  &             &  -1.0431(+4)    &   4.8853(-10)\\
%  1  &  1  &  0  & 10    &   2.1629(+5)  &             &   1.4741(+3)    &   4.3282(-10)\\
\end{tabular}
\end{ruledtabular}
\end{table}

\clearpage
\begin{figure}
\caption
  {\label{fig:v000}
Isotropic part of the quintet potential calculated at the CCSD and CCSD(T)
levels of theory. The data labeled ``RCCSD" and ``RCCSD(T)" correspond to the
uncorrected spin-restricted data, ``RCCSD-$\Delta$E" and ``RCCSD(T)-$\Delta$E"
to the size-consistency corrected data, and ``UCCSD" and ``UCCSD(T)" to the
spin-unrestricted results.
}
\end{figure}

\begin{figure}
\caption
  {\label{fig:Vfitlr}
$R$-dependent quintet potential for two selected orientations
($\theta_A,\theta_B,\phi$).  The solid lines correspond to the total fitted
potential, the dashed lines to the long-range potential, and the dotted lines
to the long-range dipole-dipole interaction.}
\end{figure}

\begin{figure}
\caption
  {\label{fig:Vfit_q4a0}
(Color) Cuts of the quintet potential (in cm$^{-1}$) for $R = 4.0$ $a_0$ and
$\phi = 0^{\circ}$. The left panel shows the fit obtained in this work and the
right panel shows the results of Ref.\ \cite{dhont:05}.}
\end{figure}

\begin{figure}
\caption
  {\label{fig:Vfit_t4a0}
(Color) Cuts of the triplet potential (in cm$^{-1}$) for $R = 4.0$ $a_0$ and
$\phi = 0^{\circ}$, calculated using Eq.\ (\ref{eq:V_CASPTn}). The upper panels
correspond to the present work and the lower panels to the work of Dhont
\textit{et al.}\ \cite{dhont:05}.} 
\end{figure}

\begin{figure}
\caption
  {\label{fig:Vfit_s4a0}
(Color) Cuts of the singlet potential (in cm$^{-1}$) for $R = 4.0$ $a_0$ and
$\phi = 0^{\circ}$, calculated using Eq.\ (\ref{eq:V_CASPTn}). The upper panels
correspond to the present work and the lower panels to the work of Dhont
\textit{et al.}\ \cite{dhont:05}.}
\end{figure}

\clearpage
\centering
\includegraphics[width=8.5cm]{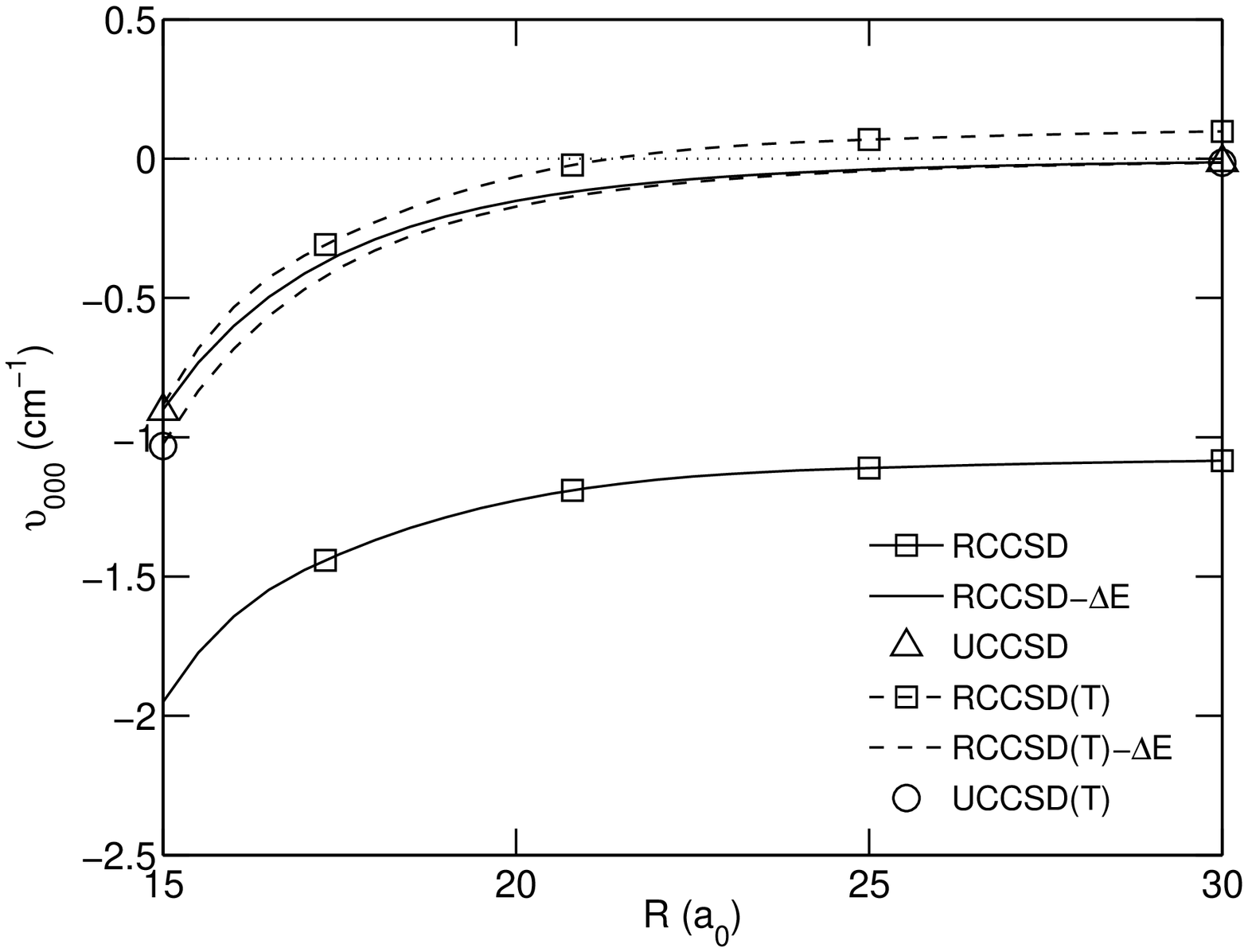}
\vfill{}
Janssen \textit{et al.}\, Fig.\ \ref{fig:v000}.

\clearpage
\centering
\includegraphics[width=8.5cm]{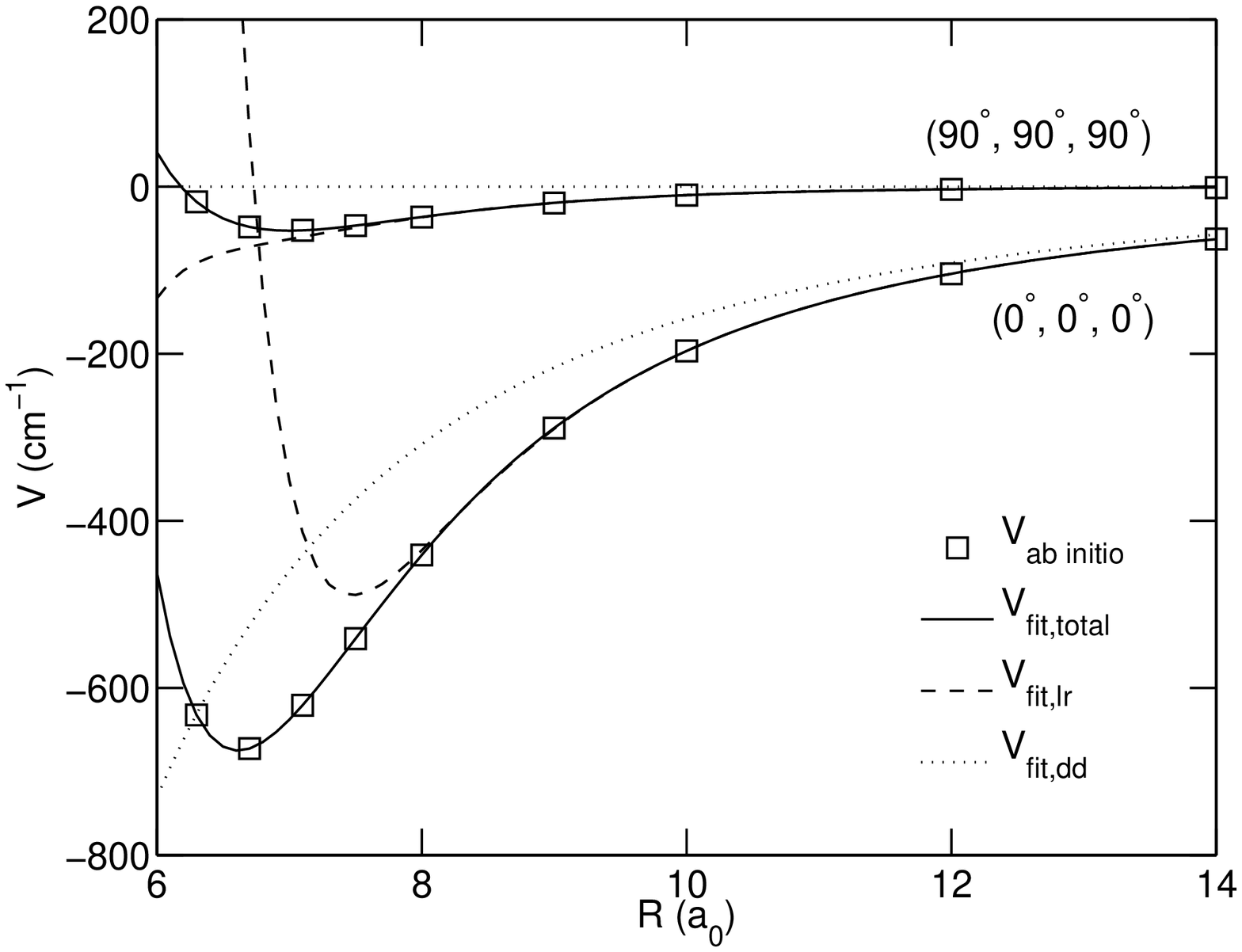}
\vfill{}
Janssen \textit{et al.}\, Fig.\ \ref{fig:Vfitlr}.

\clearpage
\centering
\includegraphics[scale=1]{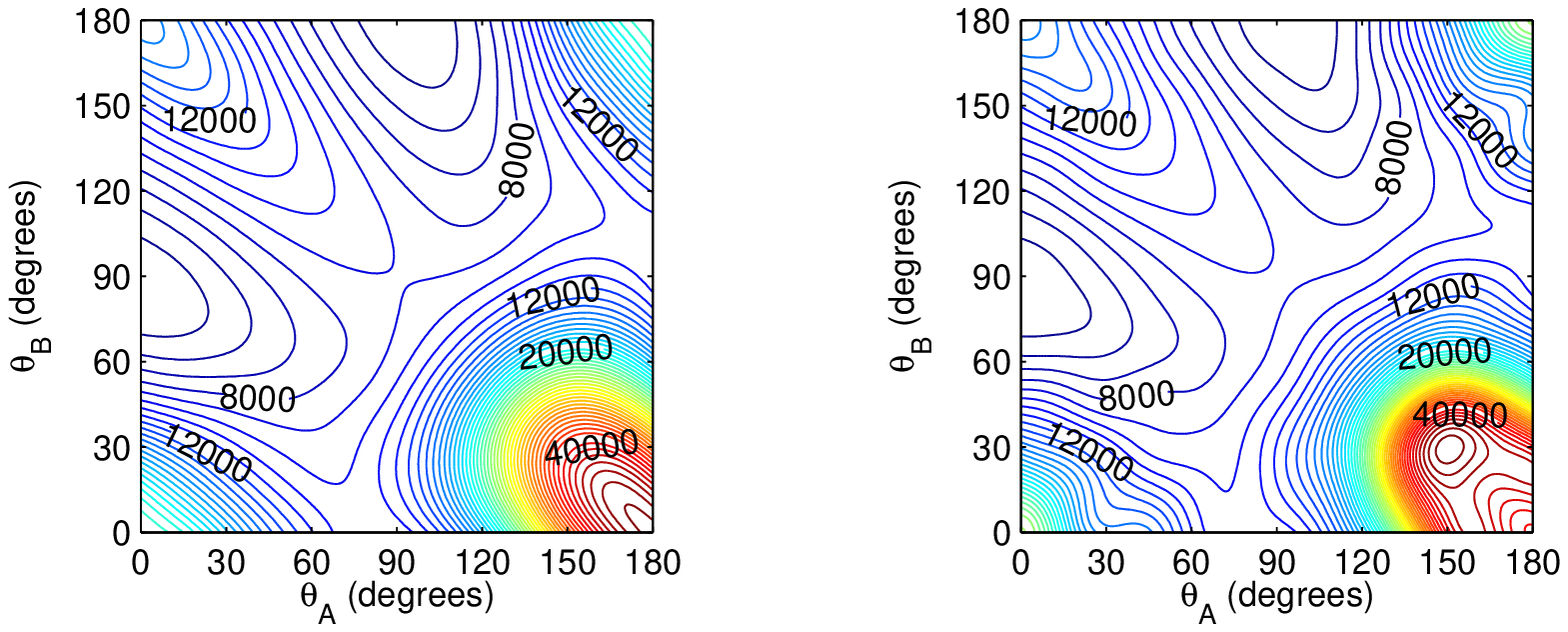}
\vfill{}
Janssen \textit{et al.}\, Fig.\ \ref{fig:Vfit_q4a0}.

\clearpage
\centering
\includegraphics[scale=1]{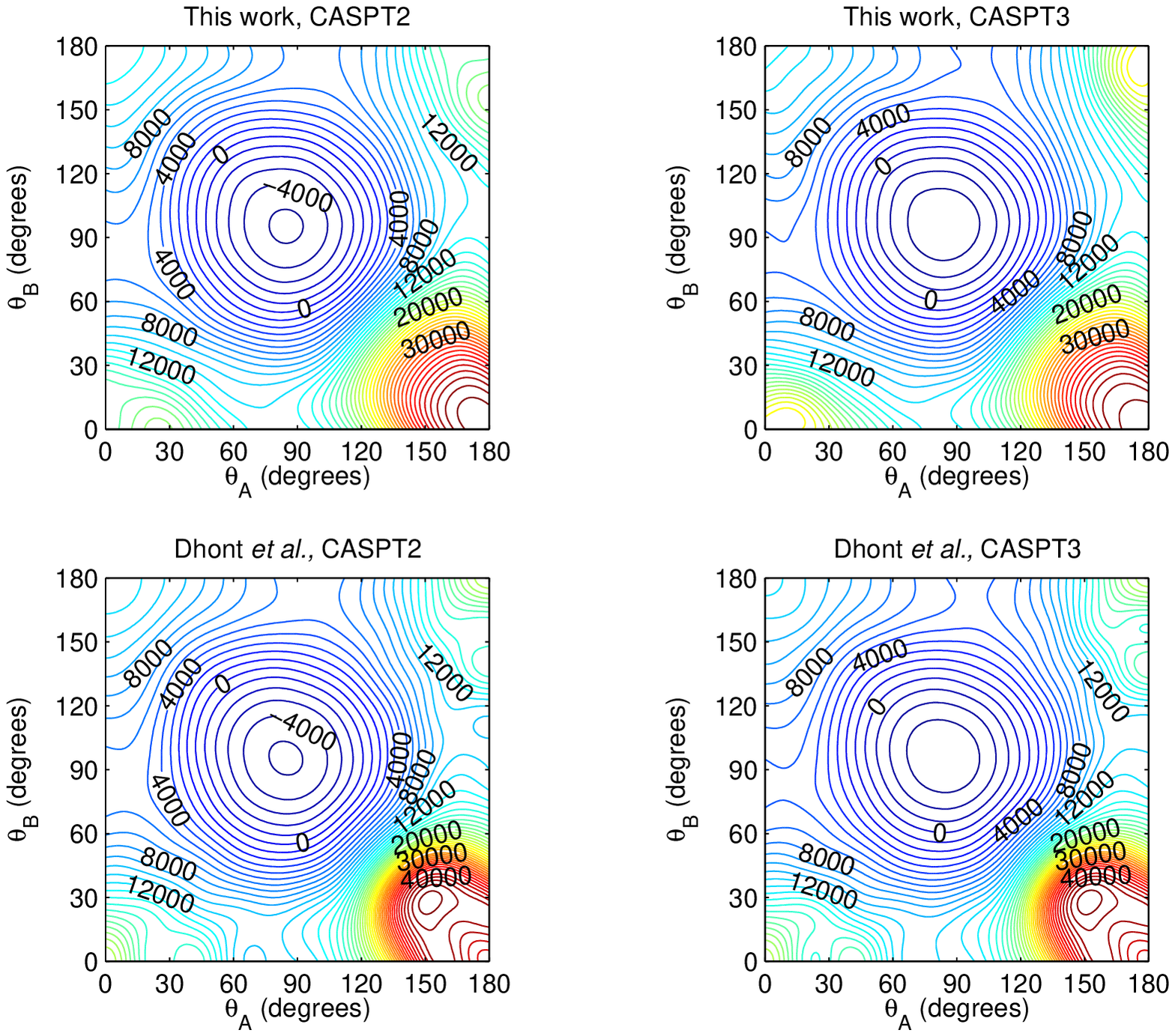}
\vfill{}  
Janssen \textit{et al.}\, Fig.\ \ref{fig:Vfit_t4a0}.

\clearpage
\centering
\includegraphics[scale=1]{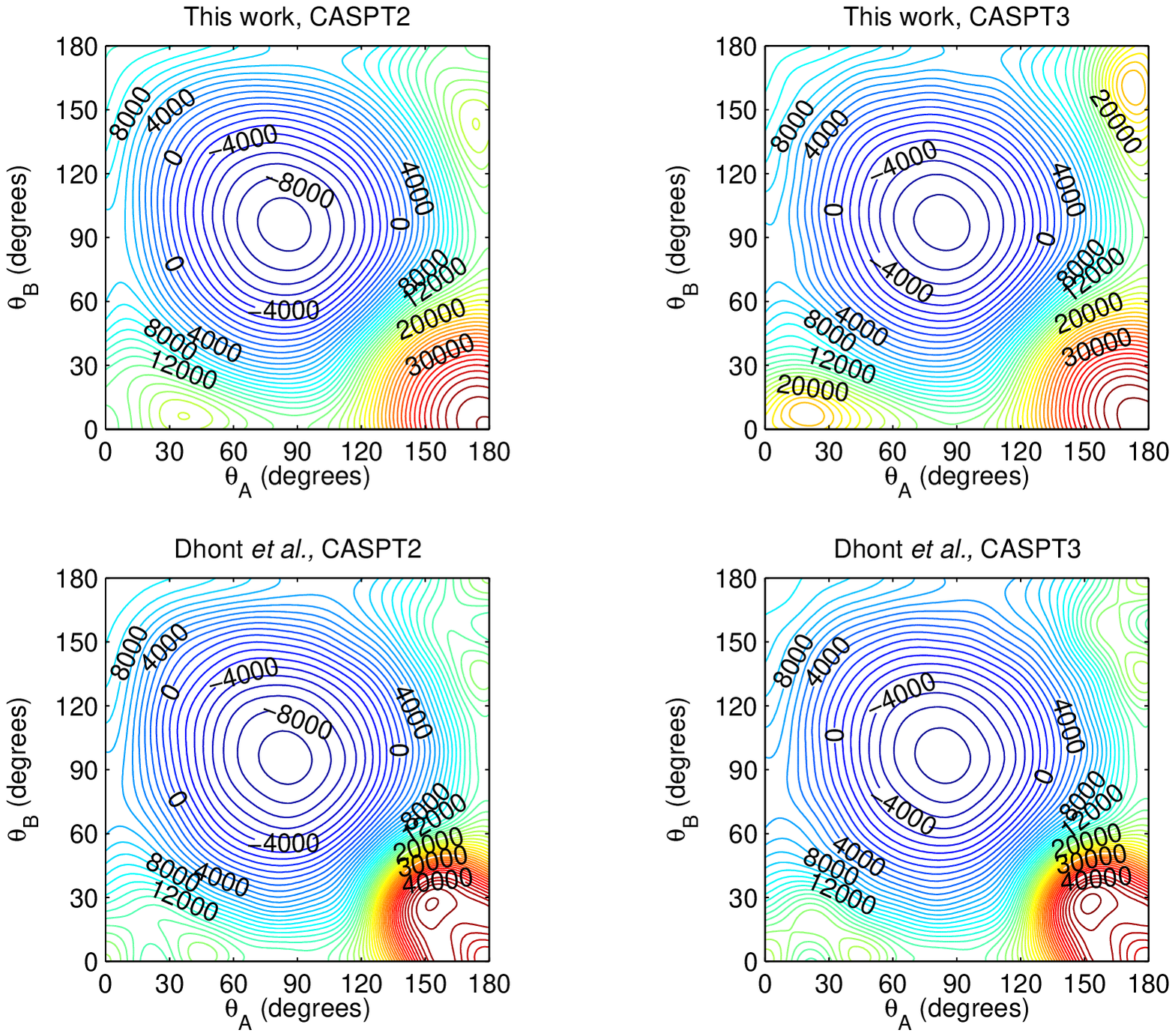}
Janssen \textit{et al.}\, Fig.\ \ref{fig:Vfit_s4a0}.

\end{document}